\documentclass[preprint2]{aastex7}

\usepackage{CJK}

\begin{document}
\begin{CJK*}{UTF8}{gbsn}

\title{Kepler's Platonic Model and Its Application to Exoplanetary Systems}

\author[0000-0002-4361-8885]{Ji Wang (王吉)}
\affiliation{Department of Astronomy, The Ohio State University, 100 W 18th Ave, Columbus, OH 43210 USA}
\email{wang.12220@osu.edu}

\begin{abstract}

Johannes Kepler's attempt to explain the arrangement of the six innermost planets of the Solar System using his \textit{Platonic Solid Model}—which postulates that planetary orbits are nested within the five Platonic solids—was ultimately unsuccessful. However, while his model failed to describe our own planetary system, Kepler was remarkably prescient in hypothesizing the existence of \textit{exoplanetary systems} that might conform to this geometric framework. In this study, we analyze all known \textit{multiple exoplanet systems} containing \textit{three to six planets} and identify those that best match the Keplerian Platonic model. Using a \textit{semi-major-axis (SMA) ratio metric} defined as the sum of squared differences between observed and theoretical semi-major-axis ratios, we find that the most well-matched \textit{three-, four-, five-, and six-planet exoplanetary systems} exhibit significantly lower discrepancy values (\(4.38141 \times 10^{-6}, 1.05799 \times 10^{-2}, 8.21728 \times 10^{-2},\) and \(2.43406 \times 10^{-1}\), respectively) compared to the inner six planets of the Solar System (\(1.26869 \times 10^{1}\)). These results demonstrate that \textit{Kepler's Platonic Model is applicable to certain exoplanetary systems}, suggesting that while the Solar System does not adhere to this idealized structure, other planetary systems may be governed by underlying geometric and mathematical principles akin to Kepler's vision. This study highlights the \textit{special nature} of these exoplanetary systems and their potential alignment with the \textit{Platonic five-element framework}.

\end{abstract}

% \keywords{\uat{Galaxies}{573} --- \uat{Cosmology}{343} --- \uat{High Energy astrophysics}{739} --- \uat{Interstellar medium}{847} --- \uat{Stellar astronomy}{1583} --- \uat{Solar physics}{1476}}


\section{Introduction} \label{sec:intro}
Johannes Kepler (1571--1630) was a pioneering astronomer whose contributions to celestial mechanics revolutionized our understanding of planetary motion \citep{kepler1609, koestler1959}. He formulated the three laws of planetary motion, which laid the foundation for Newtonian mechanics and modern orbital dynamics \citep{newton1687}. Kepler's meticulous analysis of Tycho Brahe's observational data allowed him to establish that planetary orbits are elliptical rather than circular, a breakthrough that fundamentally altered the trajectory of astronomical science \citep{field1988}. His work not only advanced heliocentric theory but also demonstrated that celestial phenomena adhere to mathematical principles, reinforcing the notion of an orderly and structured cosmos.

\begin{figure}[t]
    \centering
    \includegraphics[width=0.48\textwidth]{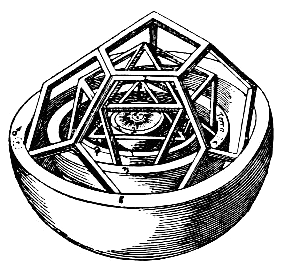}
    \caption{Kepler's Platonic solid model of the Solar System from Mysterium Cosmographicum, 1597.}
    \label{fig:kepler}
\end{figure}

Before arriving at his elliptical orbit theory, Kepler initially sought to explain the arrangement of the six then-known planets in the Solar System---Mercury, Venus, Earth, Mars, Jupiter, and Saturn---using a geometric model based on the five Platonic solids \citep{kepler1609}. This \textit{Keplerian Platonic Model} (shown in \autoref{fig:kepler}) posited that the planetary orbits were inscribed within a nested structure of the five perfect polyhedra (tetrahedron, cube, octahedron, dodecahedron, and icosahedron), with each polyhedron circumscribed by a sphere corresponding to a planetary orbit \citep{martens2000}. Kepler initially believed that this elegant geometric configuration dictated planetary distances, but the increasing accuracy of astronomical measurements ultimately revealed inconsistencies, leading him to abandon this model in favor of his more accurate elliptical laws \citep{linton2004}.

While Kepler's Platonic Solid Model failed to describe the Solar System, his vision of mathematically governed planetary structures finds new relevance in the field of exoplanet research \citep{borucki2010}. The discovery of thousands of \textit{exoplanets}---planets orbiting stars beyond our Sun---has dramatically expanded our knowledge of planetary system architectures. The \textit{Kepler Space Telescope}, launched in 2009, played a pivotal role in this transformation, detecting thousands of exoplanetary candidates through the transit method and revealing the vast diversity of planetary systems in our galaxy \citep{batalha2013}. Among these discoveries are numerous multi-planet systems with orbital configurations that differ significantly from that of the Solar System, raising the possibility that alternative mathematical frameworks, such as Kepler's Platonic model, may apply to some of these systems \citep{steffen2013}.

Despite the wealth of exoplanetary data now available, there has been no previous attempt to systematically evaluate whether any observed exoplanetary systems conform to Kepler's Platonic model. Most contemporary studies of planetary system architecture focus on empirical distributions of period ratios, planetary spacing, and resonance phenomena, without considering the possibility of underlying geometric constraints akin to those Kepler initially envisioned \citep{laskar1993, fabrycky2014}. This study aims to bridge this gap by examining whether certain exoplanetary systems exhibit structural patterns consistent with the Platonic Solid Model.

This paper is structured as follows: \textbf{Section~\ref{sec:data}} describes the data sources used for our analysis, including exoplanetary catalogs and Kepler's Platonic model configurations. \textbf{Section~\ref{sec:method}} details our methodology for identifying and quantifying system matches based on period ratio comparisons. \textbf{Section~\ref{sec:results}} presents the results of our analysis, highlighting the best-matching exoplanetary systems. Finally, \textbf{Section~\ref{sec:discussion}} discusses the implications of our findings, evaluating the extent to which Kepler's Platonic model can be applied to exoplanetary architectures and considering future directions for research in this area.

\section{Data and Processing} \label{sec:data}
The exoplanetary data used in this study were obtained from the \textit{NASA Exoplanet Archive} \citep{akeson2013}, extracted on \textit{March 2, 2025}. The dataset includes all confirmed exoplanetary systems meeting the following selection criteria: (1) the system has at least \textit{three known planets}, and (2) the system is marked as a \textit{default system} in the archive ($default\_flag = 1$). These criteria ensure that only well-characterized \textit{multi-planet systems} are included in the analysis, providing a reliable dataset for investigating potential matches to Kepler's \textit{Platonic Model}.

To prepare the data for analysis, we extract \textit{orbital period (P)} and \textit{semi-major axis (a)} values for each planet in a system. Since some entries in the dataset contain missing values for one of these parameters, we employ \textit{Kepler's Third Law} ($a^3 \sim P^2$) \citep{newton1687} to estimate the missing values, assuming a \textit{solar-mass host star} ($M_* = 1.0 M_{\odot}$). Specifically, if the semi-major axis is missing but the period is available, we compute $a$ as:

\begin{equation} \label{eq:kepler_a}
a = \left( \frac{P}{1\ yr} \right)^{\frac{2}{3}}
\end{equation}

Conversely, if the period is missing but the semi-major axis is known, we estimate $P$ as:

\begin{equation} \label{eq:kepler_p}
P = \sqrt{\left(\frac{a}{1\ AU}\right)^3}
\end{equation}

This allows us to maximize the number of usable planetary systems without discarding incomplete data.

Each multi-planet system is then \textit{individually processed} to ensure proper ordering and consistency. The planets within a system are first sorted based on their \textit{semi-major axis}, ensuring that they follow a \textit{natural outward progression from the host star}. If the semi-major axis is unavailable, the sorting is instead performed using the \textit{orbital period} as a proxy. This ensures that adjacent planets are \textit{always arranged in increasing distance from the host star}, allowing for consistent ratio computations.

Following this sorting, we compute two critical metrics for each planetary system:

\begin{enumerate}
    \item \textbf{Period Ratios}: The ratio of the orbital period of each planet relative to the next innermost planet, defined as:
    \begin{equation} \label{eq:period_ratio}
    P_{\text{ratio}, i} = \frac{P_{i+1}}{P_i}
    \end{equation}
    
    \item \textbf{Semi-Major Axis Ratios}: The ratio of the semi-major axis of each planet relative to the next innermost planet, defined as:
    \begin{equation} \label{eq:axis_ratio}
    a_{\text{ratio}, i} = \frac{a_{i+1}}{a_i}
    \end{equation}
\end{enumerate}

These ratios provide a \textit{scale-independent} way to compare planetary architectures across different star systems and are essential for matching exoplanet systems to Kepler's \textit{Platonic Solid Model}.

Finally, to facilitate comparison with Kepler's model, we filter the dataset to retain only systems with \textit{three to six planets}, ensuring that our analysis aligns with the constraints of the \textit{Platonic Solid nesting framework}. The processed dataset is then used to compute \textit{Chi$^2$ similarity metrics} between observed exoplanetary period ratios and those predicted by the \textit{Platonic Model}, enabling us to assess whether any known exoplanetary systems exhibit \textit{Keplerian geometric structuring}.

\section{The Kepler Platonic Model and Similarity Metric} \label{sec:method}

\subsection{The Kepler Platonic Model} \label{subsec:kepler_model}
Johannes Kepler's \textit{Platonic Solid Model} was an early attempt to describe planetary system architecture using the five \textit{Platonic solids}---tetrahedron, cube, octahedron, dodecahedron, and icosahedron \citep{martens2000}. Kepler hypothesized that planetary orbits were structured according to a nested arrangement of these solids, with each orbit inscribed or circumscribed within a corresponding \textit{spherical shell}. While this model ultimately failed to accurately describe the \textit{Solar System}, it proposed a fundamental idea: planetary orbits may follow a \textit{geometric pattern governed by mathematical principles}.

In the context of \textit{multi-planet exoplanetary systems}, the Kepler Platonic Model suggests that planetary spacing is determined by the \textit{geometric properties of nested polyhedral structures}. Given a sequence of \textit{N-1 nested Platonic solids}, the corresponding planetary orbit radii are defined by the \textit{outer and inner spheres} of each solid. The orbital radius of a planet associated with a given Platonic solid is dictated by either the \textit{circumscribed sphere (outer sphere)} or the \textit{inscribed sphere (inner sphere)}, leading to a set of \textit{theoretical planetary radii}.

Mathematically, the radius $R_i$ of the $i$th planetary orbit in the nested structure follows the transformation:

\begin{equation} \label{eq:radius_transformation}
R_{i+1} = R_i \times \left(\frac{R_{\text{solid}, i}}{r_{\text{solid}, i}}\right)
\end{equation}

where $R_{\text{solid}, i}$ and $r_{\text{solid}, i}$ are the \textit{outer and inner radii} of the Platonic solid defining the spacing between the $i$th and $i+1$th orbits \citep{linton2004}. The radius ratio between circumscribed and inscribed spheres for Tetrahedron, Cube, Octahedron, Dodecahedron, and Icosahedron are 3.0, 1.732, 1.732, 1.258, 1.258, respectively. These values are to be compared with SMA ratios in Table 1, 2, 3, and 4. 

\subsection{Similarity Metric: Squared Difference of Semi-Major Axis Ratios} \label{subsec:similarity_metric}
To quantitatively assess how well an exoplanetary system matches the Kepler Platonic Model, we introduce a \textit{similarity metric} based on the squared differences between the observed and theoretical \textit{semi-major axis ratios}.

For an exoplanetary system with \textit{N} planets, let the observed semi-major axis ratios be:

\begin{equation} \label{eq:observed_ratios}
(a_1, a_2, \dots, a_{N-1})
\end{equation}

and let the corresponding theoretical ratios predicted by the \textit{Kepler Platonic Model} be:

\begin{equation} \label{eq:theoretical_ratios}
(r_1, r_2, \dots, r_{N-1})
\end{equation}

We define the \textit{similarity metric} $\chi^2_{\text{semi-major}}$ as:

\begin{equation} \label{eq:chi2}
\chi^2_{\text{semi-major}} = \sum_{i=1}^{N-1} (a_i - r_i)^2
\end{equation}

This metric represents the \textit{cumulative squared deviation} between the observed and predicted semi-major axis ratios. A \textit{lower value of $\chi^2_{\text{semi-major}}$} indicates a better match between the exoplanetary system and the Kepler Platonic Model, whereas a \textit{higher value} suggests a weaker correspondence.

For each exoplanetary system in our dataset, we compute this metric for all possible \textit{nested Platonic solid configurations} and identify the \textit{best-matching theoretical model}. The resulting \textit{minimum $\chi^2_{\text{semi-major}}$} provides a quantitative measure of how closely an exoplanetary system follows the Keplerian geometric structuring.

\section{Results} \label{sec:results}
In this section, we present the results of our analysis, identifying the \textit{top-matching exoplanetary systems} that align with the \textit{Kepler Platonic Model}. The selected systems were evaluated using the \textit{semi-major axis ratio similarity metric} (see \textbf{Section~\ref{subsec:similarity_metric}}), where lower values indicate a better fit to the theoretical model. We categorize our findings by the number of planets in each system, from \textit{three to six planets}.

\subsection{Three-Planet Systems}

\begin{figure}[t]
    \centering
    \includegraphics[width=0.48\textwidth]{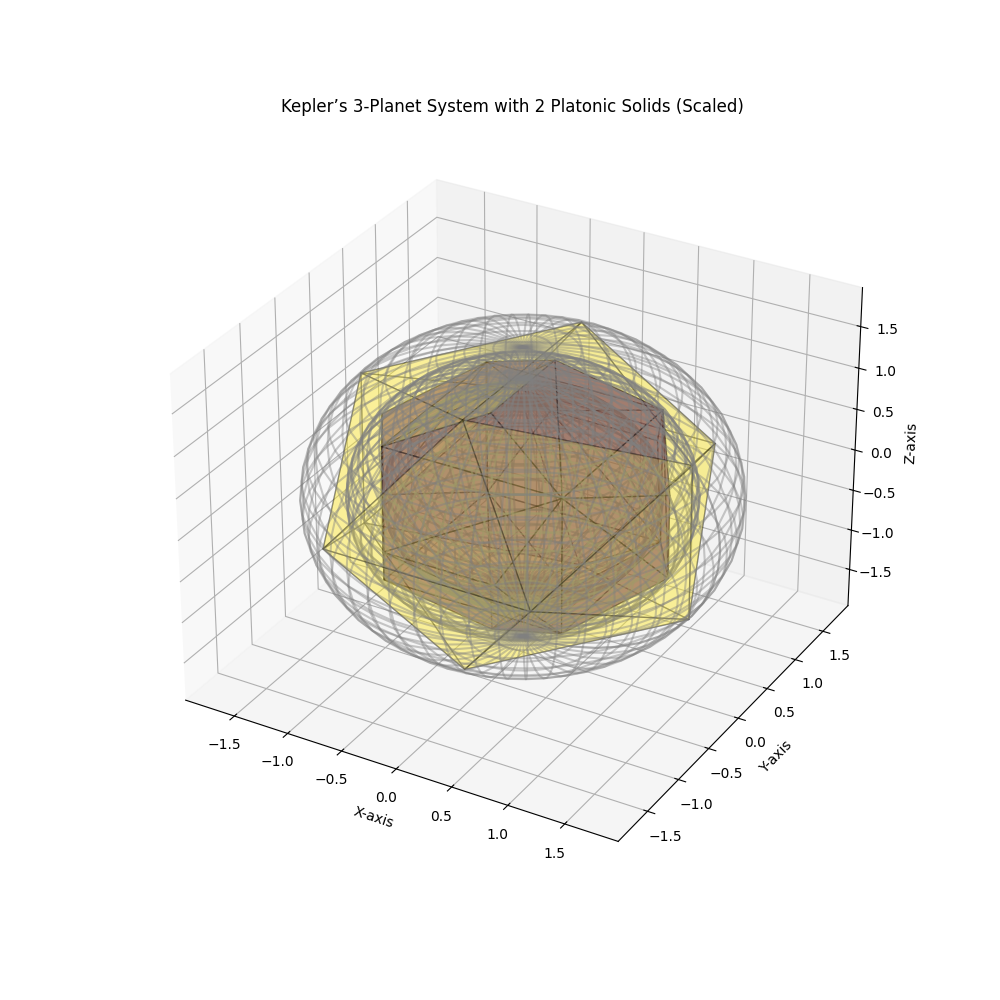}
    \caption{Kepler-271: best-matching three-planet system using the Kepler Platonic Model with a combination of a Dodecahedron and an Icosahedron~\footnote{An animation is available at: \url{https://github.com/wj198414/KeplerPlatonicModel/blob/00e463f25bb6ea4302412f591171f1aa6e53aab5/kepler_animation.gif}}.}
    \label{fig:3_planets}
\end{figure}

Among the \textit{three-planet systems} analyzed, the best matches to the \textit{Kepler Platonic Model} are \textit{Kepler-271, Teegarden's Star, and K2-198} (\autoref{tab:3planet}). These systems exhibit \textit{period ratios} closely aligned with the theoretically predicted values, with \textit{Kepler-271} showing an exceptionally low $\chi^2$ value of $4.38 \times 10^{-6}$ (\autoref{fig:3_planets}). These results suggest that \textit{Kepler-271}, in particular, exhibits a planetary spacing structure remarkably similar to that predicted by the nested Platonic solids framework.

\begin{table*}[t]
\centering
\caption{Best-Matching Three-Planet Systems}
\begin{tabular}{lccc}
\hline
\textbf{System} & \textbf{Periods (days)} & \textbf{SMA Ratios} & $\chi^2$ \\
\hline
Kepler-271 & [5.25, 7.41, 10.44] & [1.258, 1.256] & $4.38 \times 10^{-6}$ \\
Teegarden's Star & [4.91, 11.42, 26.13] & [1.757, 1.738] & $6.48 \times 10^{-4}$ \\
K2-198 & [3.36, 7.45, 17.04] & [1.701, 1.736] & $1.01 \times 10^{-3}$ \\
\hline
\end{tabular}
\label{tab:3planet}
\end{table*}

\subsection{Four-Planet Systems}

\begin{figure}[t]
    \centering
    \includegraphics[width=0.48\textwidth]{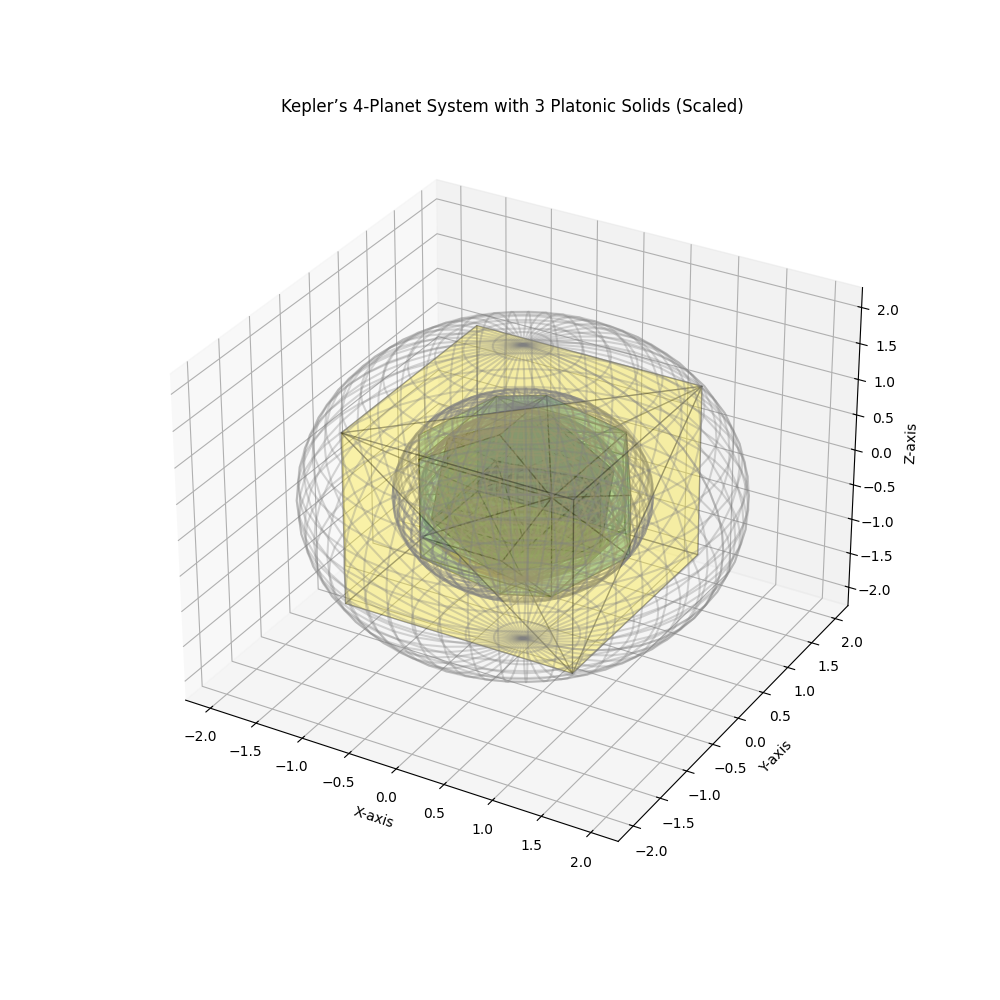}
    \caption{HD 215152: best-matching four-planet system using the Kepler Platonic Model with a combination of an Icosahedron, a Dodecahedron, and Cube.}
    \label{fig:4_planets}
\end{figure}

For \textit{four-planet systems}, the top matches include \textit{HD 215152, K2-133, and K2-32} (\autoref{tab:4planet}). The best fit, \textit{HD 215152}, has a $\chi^2$ value of $1.06 \times 10^{-2}$ (\autoref{fig:4_planets}). These systems demonstrate that four-planet exoplanetary configurations can sometimes conform well to the geometric constraints imposed by Kepler's nested Platonic solids model.

\begin{table*}[t]
\centering
\caption{Best-Matching Four-Planet Systems}
\begin{tabular}{lccc}
\hline
\textbf{System} & \textbf{Periods (days)} & \textbf{SMA Ratios} & $\chi^2$ \\
\hline
HD 215152 & [5.76, 7.28, 10.86, 25.20] & [1.169, 1.306, 1.752] & $1.06 \times 10^{-2}$ \\
K2-133 & [3.07, 4.87, 11.02, 26.58] & [1.359, 1.725, 1.798] & $1.45 \times 10^{-2}$ \\
K2-32 & [4.35, 8.99, 20.66, 31.72] & [1.623, 1.741, 1.331] & $1.73 \times 10^{-2}$ \\
\hline
\end{tabular}
\label{tab:4planet}
\end{table*}

\subsection{Five-Planet Systems}

\begin{figure}[h]
    \centering
    \includegraphics[width=0.48\textwidth]{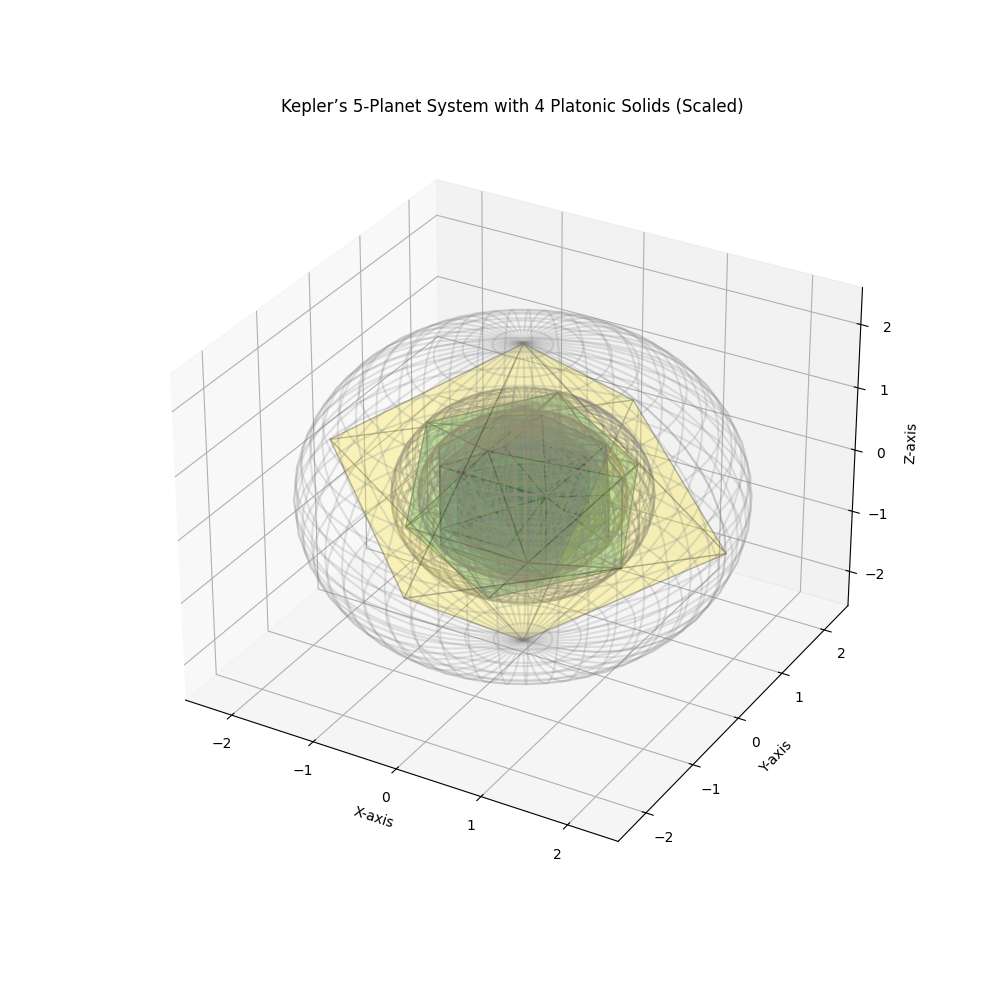}
    \caption{K2-268: best-matching five-planet system using the Kepler Platonic Model with a combination of a Cube, a Dodecahedron, an Icosahedron', and an Octahedron.}
    \label{fig:5_planets}
\end{figure}

For \textit{five-planet systems}, the best matches were found in \textit{K2-268, Kepler-122, and Kepler-33} (\autoref{tab:5planet}), with \textit{K2-268} yielding the lowest $\chi^2$ value of $8.22 \times 10^{-2}$ (\autoref{fig:5_planets}). Although these five-planet systems do not achieve the same level of agreement as three- and four-planet cases, their orbital structures still display a notable alignment with the Platonic Solid Model.

\begin{table*}[t]
\centering
\caption{Best-Matching Five-Planet Systems}
\begin{tabular}{lccc}
\hline
\textbf{System} & \textbf{Periods (days)} & \textbf{SMA Ratios} & $\chi^2$ \\
\hline
K2-268 & [2.15, 4.53, 6.13, 9.33, 26.27] & [1.642, 1.224, 1.323, 1.994] & $8.22 \times 10^{-2}$ \\
Kepler-122 & [5.77, 12.47, 21.59, 37.99, 56.27] & [1.672, 1.442, 1.458, 1.299] & $1.14 \times 10^{-1}$ \\
Kepler-33 & [5.67, 13.18, 21.78, 31.78, 41.03] & [1.756, 1.398, 1.286, 1.186] & $1.18 \times 10^{-1}$ \\
\hline
\end{tabular}
\label{tab:5planet}
\end{table*}

\begin{figure}[h]
    \centering
    \includegraphics[width=0.48\textwidth]{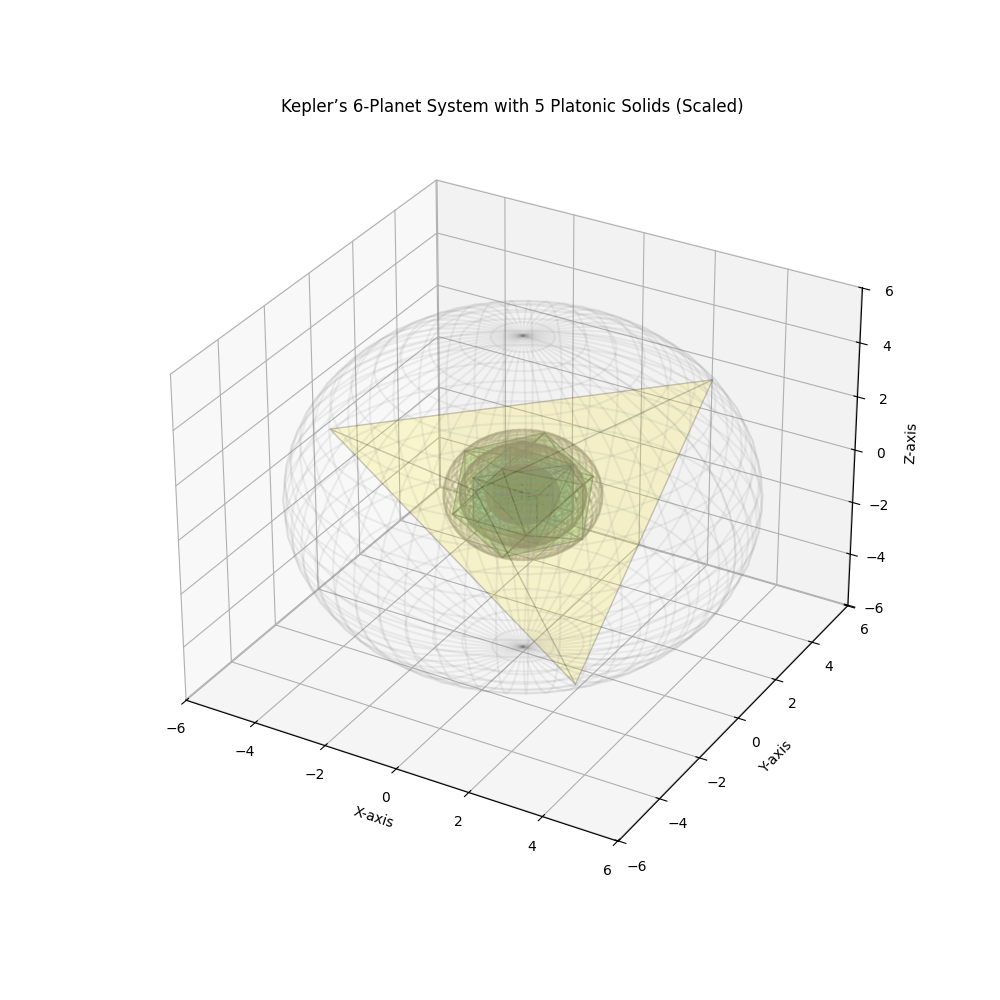}
    \caption{HD 34445: best-matching six-planet system using the Kepler Platonic Model with a combination of an Octahedron, a Dodecahedrona, a Cube, an Icosahedron, and a Tetrahedron.}
    \label{fig:6_planets}
\end{figure}

\subsection{Six-Planet Systems}

For \textit{six-planet systems}, the top three matches include \textit{HD 34445, K2-138, and Kepler-80} (\autoref{tab:6planet}). Among these, \textit{HD 34445} exhibits the best agreement with a $\chi^2$ value of $2.43 \times 10^{-1}$ (\autoref{fig:6_planets}). The six-planet configurations show a higher degree of deviation compared to systems with fewer planets. However, the fact that some systems still show notable alignment with the Platonic Model suggests that this geometric framework may be applicable to a subset of exoplanetary architectures.

\begin{table*}[t]
\centering
\caption{Best-Matching Six-Planet Systems}
\begin{tabular}{lccc}
\hline
\textbf{System} & \textbf{Periods (days)} & \textbf{SMA Ratios} & $\chi^2$ \\
\hline
HD 34445 & [49.18, 117.87, 214.67, 676.8, 1049.0, 5700.0] & [1.793, 1.491, 2.149, 1.342, 3.072] & $2.43 \times 10^{-1}$ \\
K2-138 & [2.35, 3.56, 5.40, 8.26, 12.76, 41.97] & [1.318, 1.321, 1.327, 1.336, 2.212] & $9.49 \times 10^{-1}$ \\
Kepler-80 & [0.99, 3.07, 4.64, 7.05, 9.52, 14.65] & [2.132, 1.317, 1.321, 1.222, 1.332] & $1.09 \times 10^{0}$ \\
\hline
\end{tabular}
\label{tab:6planet}
\end{table*}

\subsection{Summary of Results}

These results demonstrate that \textit{several exoplanetary systems exhibit orbital spacing structures that align well with the Kepler Platonic Model}. The \textit{three- and four-planet systems} yield the \textit{best matches}, while \textit{five- and six-planet systems} show progressively higher deviations. Notably, all of these exoplanetary systems exhibit a \textit{closer match to the Platonic Model than the inner six planets of the Solar System}, reinforcing the hypothesis that Kepler’s geometric framework may be relevant to certain planetary architectures beyond our own system.

\section{Discussion} \label{sec:discussion}

\subsection{The Solar System in the Context of Exoplanetary Architectures} \label{subsec:solarsystem}
One of the most striking findings from this study is that the \textit{inner six planets of the Solar System do not conform to Kepler's Platonic Model}, whereas certain exoplanetary systems exhibit strong alignment with this geometric framework. The \textit{semi-major axis ratio similarity metric} for the Solar System's inner six planets is $1.26869 \times 10^1$—\textit{two orders of magnitude larger} than the best-matching \textit{three-, four-, five-, and six-planet exoplanetary systems}, which yielded values of $4.38141 \times 10^{-6}, 1.05799 \times 10^{-2}, 8.21728 \times 10^{-2},$ and $2.43406 \times 10^{-1}$, respectively. This substantial deviation suggests that while the \textit{Solar System lacks a strict geometric structuring}, other exoplanetary systems may adhere more closely to the principles outlined in Kepler's \textit{Platonic Solid Model} \citep{laskar1993, fabrycky2014}.

This result aligns with broader trends in \textit{exoplanetary science}, where the diversity of planetary system architectures has challenged previous assumptions about the \textit{Solar System as a universal template} for planetary formation \citep{morton2014}. The discovery of compact multi-planet systems, resonant chain configurations, and super-Earth populations demonstrates that planetary architectures are highly varied, possibly shaped by \textit{initial formation conditions, disk properties, and dynamical evolution} \citep{raymond2020}. The fact that \textit{some} exoplanetary systems exhibit geometric structuring suggests that specific conditions—yet to be fully understood—may favor the emergence of \textit{Platonic-like orbital spacing}.

\begin{figure}[t]
    \centering
    \includegraphics[width=0.48\textwidth]{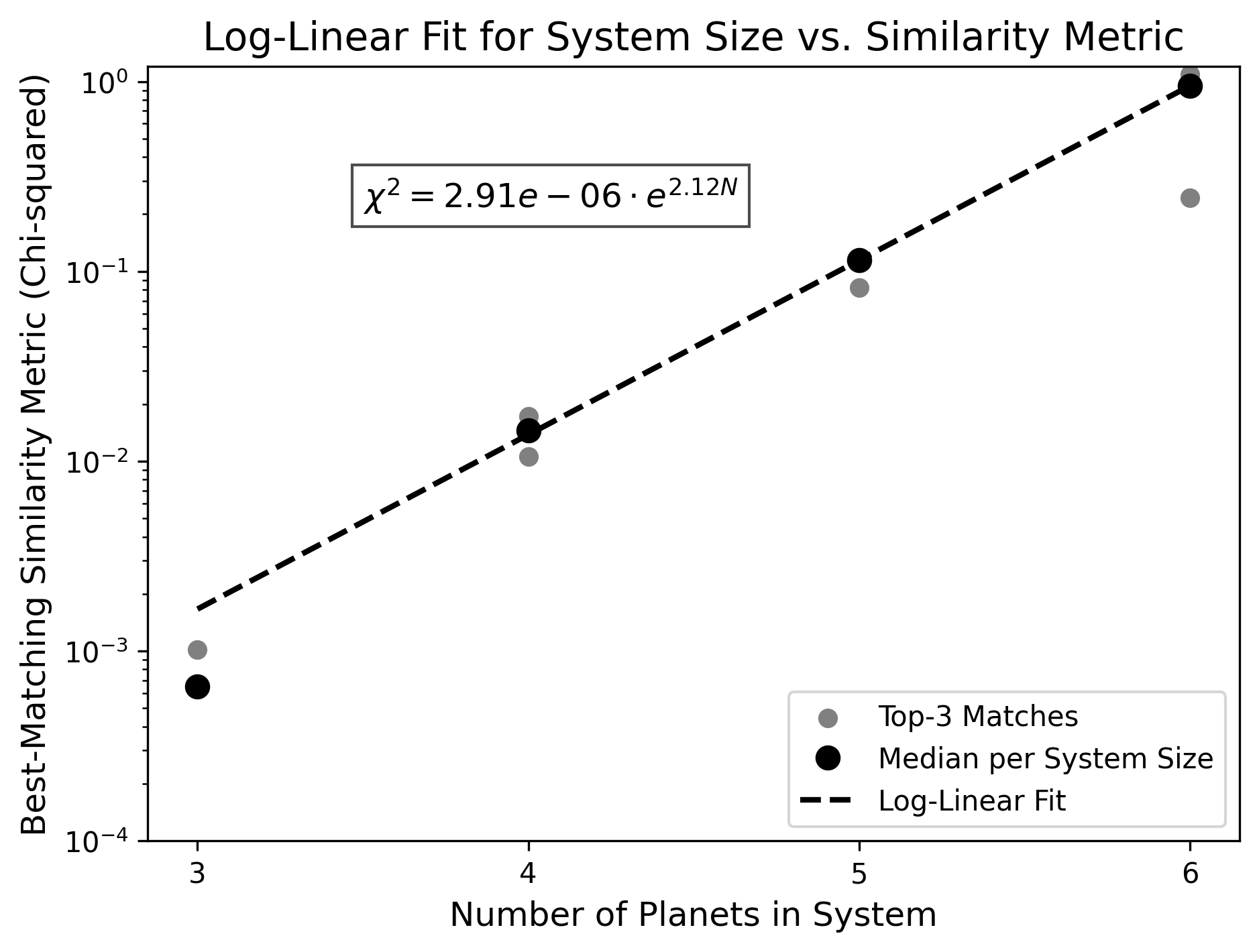}
    \caption{Log-linear relationship between the number of planets in a system and the best-matching similarity metric. The black dashed line represents the best-fit log-linear trend, while black markers indicate the median per system size.}
    \label{fig:log_linear}
\end{figure}

Furthermore, the relationship between the \textit{number of planets in a system and the best-matching similarity metric} appears to \textit{follow a log-linear trend}, which is illustrated in \autoref{fig:log_linear}. This trend implies that as planetary systems grow more complex, deviations from perfect geometric structures increase, potentially due to interactions between planets, migration processes, or external perturbations. This observation supports the idea that while \textit{idealized geometric principles may govern certain planetary configurations}, real-world planetary systems are still subject to \textit{chaotic dynamics and environmental influences} that can alter their final structure \citep{chambers1998}.

\subsection{The Special Nature of Platonic-Aligned Exoplanetary Systems} \label{subsec:platonic_systems}
The planetary systems identified in this study as the \textit{best matches} to the Kepler Platonic Model may represent a \textit{unique subclass of exoplanetary architectures}, potentially governed by deeper \textit{geometric or physical principles}. One possible explanation for their orbital configurations is their alignment with the \textit{Platonic Five-Element Framework}, an ancient philosophical concept that assigns the \textit{five Platonic solids} to the classical elements \citep{plato2008}:

\begin{itemize}
    \item \textbf{Tetrahedron} $\rightarrow$ Fire
    \item \textbf{Cube} $\rightarrow$ Earth
    \item \textbf{Octahedron} $\rightarrow$ Air
    \item \textbf{Dodecahedron} $\rightarrow$ Aether (Cosmos)
    \item \textbf{Icosahedron} $\rightarrow$ Water
\end{itemize}

In the context of planetary formation, this framework suggests that planetary orbits may be \textit{influenced by fundamental physical symmetries} that arise during system formation. If certain planetary systems form within disks that exhibit \textit{geometrically self-organized structures}, they may naturally evolve into \textit{nested configurations corresponding to Platonic solids} \citep{laskar2017}. Alternatively, \textit{resonant interactions} between planets, or \textit{early-stage migration patterns}, could impose constraints that favor spacing ratios near \textit{those predicted by Kepler's model} \citep{batygin2020}.

While the evidence presented in this study supports the \textit{existence of Platonic-aligned planetary systems}, additional data is required to confirm this hypothesis. Future surveys—particularly those from \textit{next-generation missions like PLATO, JWST, and the Roman Space Telescope}—could provide deeper insights into the frequency and physical properties of these special planetary systems \citep{barclay2018}. A more comprehensive statistical analysis incorporating \textit{orbital resonances, planetary masses, and system stability criteria} will be essential for understanding whether these configurations arise due to \textit{underlying physical laws or are the result of stochastic formation processes}.

Ultimately, this study suggests that while \textit{Kepler's Platonic Model is not universally applicable}, it may describe a \textit{specific class of exoplanetary systems} where planetary spacing is dictated by \textit{fundamental geometric principles} rather than \textit{purely stochastic or dynamically chaotic processes}. If confirmed, these systems would represent a remarkable case where \textit{astronomical reality aligns with the mathematical elegance envisioned by Kepler over four centuries ago}.

%% Please use the acknowledgment and contribution environments. This will 
%% be anonomyized when the "anonymous" style option is used. 
\begin{acknowledgments}
The bulk of the work in this manuscript is done by ChatGPT and deepseek. The work is inspired by a LACMA trip summary by David Weinberg during the famous OSU coffee on Feb 28, 2025. 
\end{acknowledgments}

\bibliography{sample7}{}
\bibliographystyle{aasjournal}



\end{CJK*}
\end{document}